\documentclass[aps,twocolumn,prd,superscriptaddress,showpacs,nofootinbib]{revtex4-1}

\usepackage[utf8]{inputenc}

\usepackage{diagbox}

\usepackage{mathtools}
\usepackage{amsfonts}
\usepackage{mathrsfs}
\usepackage{bbm}
\usepackage{slashed}

\usepackage{graphicx}
\usepackage{color}
\usepackage{array}

\usepackage{placeins}
\usepackage{booktabs}
\usepackage{makecell}
\usepackage{epstopdf}
\usepackage[caption=false]{subfig}

\usepackage{xspace}
\usepackage{siunitx}
\usepackage{hyperref}
\usepackage[nameinlink]{cleveref}
\usepackage{appendix}

\usepackage{xifthen}
\usepackage{xcolor}
\hypersetup{
	colorlinks,
	linkcolor={red!75!black},
	citecolor={blue!75!black},
	urlcolor={blue!75!black}
}

\setkeys{Gin}{width=0.48\textwidth}

\graphicspath{
	{./figures/}
	{../figures/}
}

\def\pslash{p\llap{/}}
\def\qslash{q\llap{/}}
\def\kslash{k\llap{/}}

\def\s0#1#2{\mbox{\small{$ \frac{#1}{#2} $}}}
\def\0#1#2{\frac{#1}{#2}}

\newcommand{\ltpre}[1][0pt]{\mathrel{\raisebox{#1}{\scriptsize$\parallel$}}}
\newcommand{\lt}{{\ltpre[1.2pt]}}
\newcommand{\trans}{ \bot}

\def\eq#1{\eqref{#1}}

\def\fig#1{Figure \ref{#1}}
\def\Fig#1{Figure~\ref{#1}}

\newcommand{\tr}{{\text{tr}}}

\newcommand{\sumint}{\int\hspace{-4.8mm}\sum}
\newcommand{\imag}{\text{i}}

\newcommand{\gettitle}{Chiral phase structure and critical end point in QCD}

\hypersetup{
	pdftitle={\gettitle},
	pdfauthor={Gao, Pawlowski},
	pdfkeywords={functional methods}
		{correlations functions} {phase structure}
		{functional renormalisation group} {Dyson-Schwinger equation},
	bookmarksopen=true,
	bookmarksopenlevel=2,
	bookmarksnumbered=true
}

\begin{document}

\title{\gettitle}

\author{Fei Gao}
\affiliation{Institut f{\"u}r Theoretische Physik,
	Universit{\"a}t Heidelberg, Philosophenweg 16,
	69120 Heidelberg, Germany
}

\author{Jan M. Pawlowski}
\affiliation{Institut f{\"u}r Theoretische Physik,
	Universit{\"a}t Heidelberg, Philosophenweg 16,
	69120 Heidelberg, Germany
}
\affiliation{ExtreMe Matter Institute EMMI,
	GSI, Planckstr. 1,
	64291 Darmstadt, Germany
}


\begin{abstract}
	We map out the QCD phase structure at finite temperature and chemical potential for 2-flavour and 2 + 1-flavour QCD. This is done within a generalised functional approach to QCD put forward in \cite{Gao:2020qsj}. Specifically we compute the quark propagator and the finite-temperature and density fluctuations of the gluon propagator and the quark-gluon vertex on the basis of precision data for vacuum correlation functions. The novel ingredient is the direct self-consistent computation of the DSEs for the dressings of the quark-gluon vertex, in contrast to the common use of STI-inspired vertices.
	
	For small densities the results for the chiral order parameter agree with the respective lattice and functional renormalisation group results, for large densities the present results are in a quantitative agreement with the latter, including the location of the critical end point.
\end{abstract}

\maketitle

\section{Introduction}\label{sec:Introduction}

The QCD phase structure at finite density has been the subject of many works within functional methods in the past two decades, for results with the functional renormalisation group (fRG) see e.g.~\cite{Fu:2018qsk, Fu:2019hdw, Leonhardt:2019fua, Braun:2019aow, Braun:2020ada, Dupuis:2020fhh}, for results with Dyson-Schwinger equations (DSE) see e.g.~\cite{Roberts:2000aa, Fischer:2014ata, Gao:2016qkh, Gao:2020qsj, Fischer:2018sdj, Isserstedt:2019pgx, Gao:2020qsj}. At small densities these studies are accompanied by respective  lattice studies, see e.g.~\cite{Bazavov:2017dus, Bazavov:2017tot, Bonati:2018nut, Borsanyi:2018grb, Bazavov:2018mes, Guenther:2018flo, Ding:2019prx, Borsanyi:2020fev}. By now the results from both, lattice and functional methods, agree at small densities. In turn, at larger densities the lattice is hampered by the sign problem, while the approximations to the full QCD effective action within functional approaches require systematic qualitative improvements. This is the topic of the present work, bases on the generalised functional approach (fRG-DSE) put forward in \cite{Gao:2020qsj}. These advances should finally lead to a quantitative theoretical access to the QCD phase structure, mapped out by running and planned heavy-ion experiments, for reviews see \cite{Luo:2017faz, Adamczyk:2017iwn, Andronic:2017pug, Stephanov:2007fk, Andersen:2014xxa,  Shuryak:2014zxa, Pawlowski:2014aha, Roberts:2000aa, Fischer:2018sdj, Yin:2018ejt}.

In the generalised functional approach put forward in \cite{Gao:2020qsj}, the DSEs of correlation functions at finite temperature and chemical potential have been solved in an expansion about quantitative $N_f=2$ flavour QCD vacuum correlation functions obtained with the fRG in \cite{Cyrol:2017ewj}. Special emphasis has been put on the full quark-gluon vertex, which is the key input for the solution of the quark gap equation. In \cite{Gao:2020qsj} the thermal and density fluctuations of the quark gluon vertex have been derived from the respective Slavnov-Taylor  identities (STIs), further gauge symmetry constraints, and the fRG vacuum vertex. The results for the phase boundary for small chemical potential agree well with the fRG, \cite{Fu:2019hdw} and lattice results, \cite{Borsanyi:2020fev} (WB), \cite{Bazavov:2018mes} (HotQCD). For larger chemical potential the results from functional methods, \cite{Gao:2020qsj, Fu:2019hdw}, still agree, while there are no lattice results. However, the locations of the critical end point (CEP) for fRG- and DSE-approaches differ, even though they are still compatible with in the rapidly systematic error of both computations in this regime, see the respective discussions in \cite{Gao:2020qsj, Fu:2019hdw}.

In the present work compute the QCD phase structure within systematic qualitative improvements of the truncation level used in \cite{Gao:2020qsj}. In particular, the reliability of the results at larger chemical potential is enhanced qualitatively: the novel ingredient is the self-consistent solution of the DSEs for strange, thermal and density fluctuations of the quark-gluon vertex in contradistinction to the standard STI-vertex constructions.

This first principles approach gives access to the phase structure of QCD without the need of phenomenological infrared parameters. The only input are the fundamental parameters in QCD: the strong coupling and the current quark masses. At small densities the chiral phase boundary computed in the present work agrees quantitatively with that from functional studies, \cite{Fu:2019hdw, Gao:2020qsj} and lattice data, e.g.~\cite{Borsanyi:2020fev, Bazavov:2018mes}. At larger densities the present  result agrees quantitatively with the results from the fRG-study in \cite{Fu:2019hdw}, including the location of the CEP. This is a chiefly important reliability test of the respective results, in particular given the different resummation schemes implemented for the set of fRG- and DSE-equations.

\section{fRG-assisted DSEs}\label{sec:FRG-DSE}

\begin{figure*}
 \subfloat[DSE for the quark and gluon propagators.\hfill\textcolor{white}{.}]{
\includegraphics[width=0.48\textwidth]{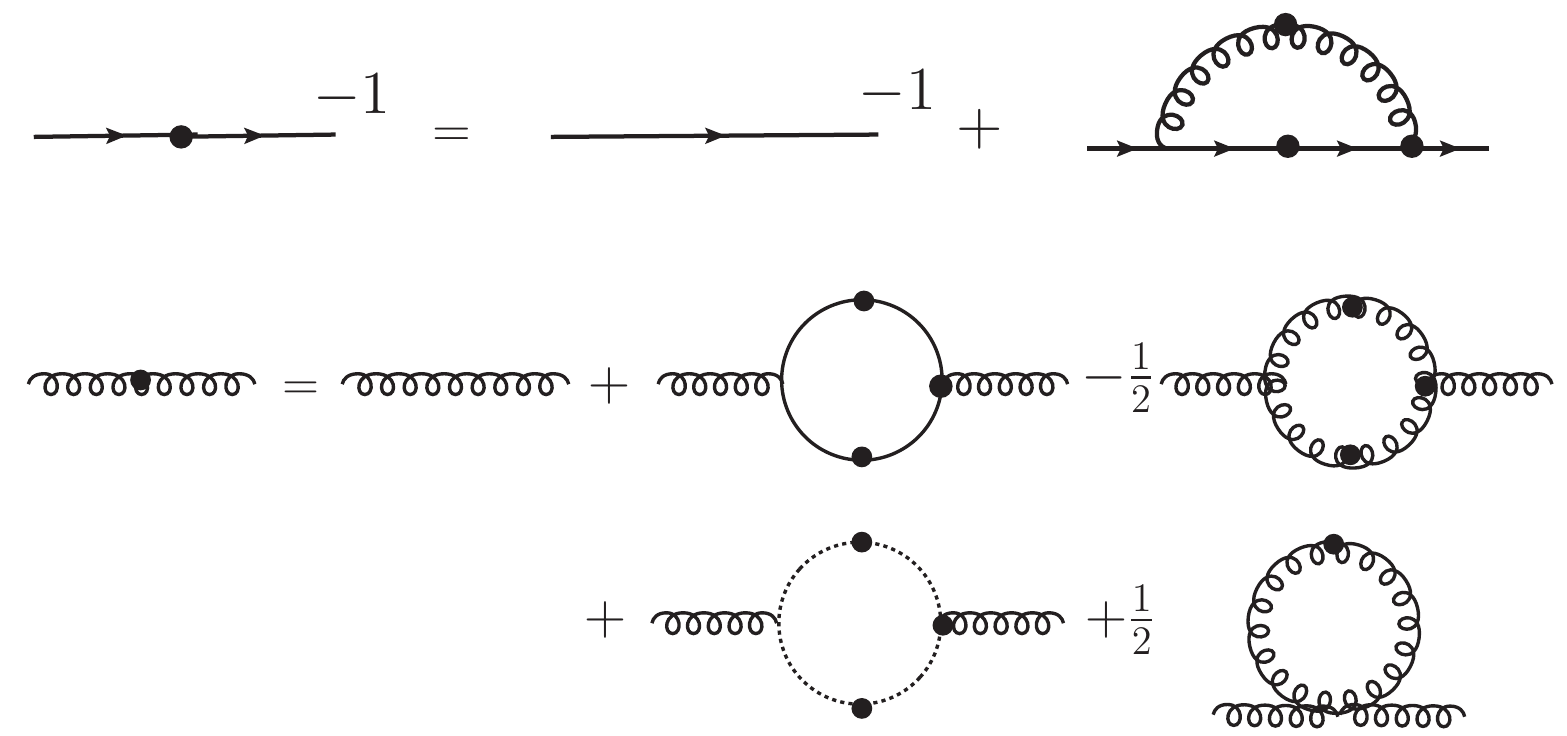}
    \label{fig:gap}} \hfill   \subfloat[\quad DSE for the quark gluon vertex.\hfill\textcolor{white}{.}]{\includegraphics[width=0.5
\textwidth]{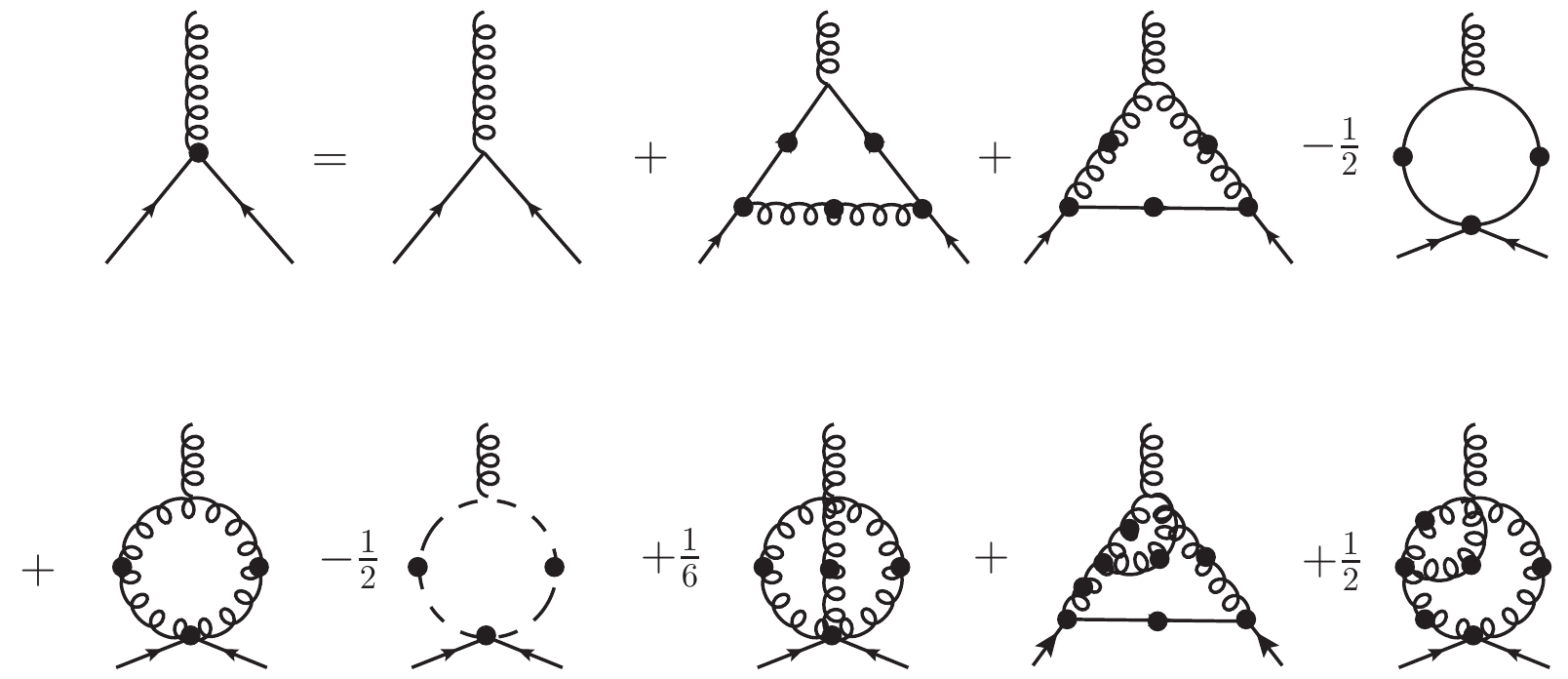}
\label{fig:Vertices}}
\caption{Spiralling lines represent gluons, solid lines quarks, and black blobs are dressed vertices, grey is the classical vertex. In the present work the quark propagator is obtained from its gap equation depicted above. In turn, we solve the gluon gap equation and DSE of quark gluon vertex for the strange-, thermal and density fluctuations, $\Delta\Gamma^{(2)}_{AA}$ and $\Delta\Gamma^{(2)}_{q\bar q A}$, as described in \eq{eq:Gn} and below.  A discussion of the reliability of this approximation is found in the text. \hfill\textcolor{white}{.} }\label{fig:main_result}
\end{figure*}
We first outline the generalised  functional approach put forward and detailed in \cite{Gao:2020qsj}. Then the improved truncation used in the present work is discussed.

\subsection{Setup for fRG-assisted DSEs}

The DSEs for correlation functions are one- or two-loop exact loop relations that rely on full propagators as well as full and classical vertices. They have to be renormalised non-perturbatively in the vacuum, and finite temperature and density fluctuations do not change the renormalisation.
This suggests to parameterise the full correlation functions as
\begin{align} \label{eq:Gn}
	\left. \Gamma^{(n)}_{}(p) \right|_{T,\mu_B, N_f}  =
	\left. \Gamma^{(n)}_{}(p)\right|_{0,0,2} +
	\Delta\Gamma^{(n)}_{}(p)\,.
\end{align}
In \eq{eq:Gn}, $\Gamma^{(n)}$ is a general (1PI) $n$-point function in $N_f$-flavour QCD at finite temperature and density, which is expanded about its $N_f=2$ vacuum counter-part with isospin symmetry. In the present work and in \cite{Gao:2020qsj}, the $N_f=2$ vacuum QCD data have been taken from the quantitative fRG-study \cite{Cyrol:2017ewj}. The two-flavour vacuum QCD expansion point can be easily changed to other ones from either fRG, DSE or the lattice, subject to the existence of respective quantitative data sets. For functional results, see e.g.\ \cite{Eichmann:2014xya,  Williams:2014iea, Mitter:2014wpa, Fischer:2014ata, Eichmann:2015kfa, Cyrol:2016tym, Cyrol:2017ewj, Cyrol:2017qkl, Aguilar:2016lbe, Aguilar:2017dco, Aguilar:2018epe, Reinosa:2015oua, Reinosa:2016iml, Maelger:2018vow, Maelger:2019cbk}, for lattice results, see e.g.\ \cite{Bowman:2005vx, Fischer:2010fx, Maas:2011ez, Maas:2011se, Aouane:2012bk, Silva:2013maa, Athenodorou:2016oyh, Sternbeck:2017ntv, Oliveira:2018lln, Oliveira:2018lln, Boucaud:2018xup, Leutnant:2018dry, Zafeiropoulos:2019flq, Sternbeck:2019twy, Aguilar:2019uob}. Importantly, the thermal and density fluctuations in the $\Delta\Gamma^{(n)}$ do not require renormalisation: the respective loop integrals are finite.

\subsection{Truncation scheme} \label{sec:truncation}

In the present work we use the split \eq{eq:Gn} for the gluon propagator and the quark-gluon vertex, and solve the respective DSEs for $\Delta\Gamma^{(2)}_{AA}$ and $\Delta\Gamma^{(3)}_{q\bar qA}$. In the latter case  the DSE is only solved for the dominant three tensor structures, while the others five tensor structures are determined from symmetry constraints. The quantitative reliability of this procedure is discussed in detail and tested in \cite{Gao:2020qsj}. Further information about this hierarchy as well as the derivation of the symmetry constraints are found in the fRG-works \cite{Cyrol:2017ewj, Mitter:2014wpa}. Moreover, for the strange-quark--gluon vertex we use
\begin{align}\label{eq:strangequark}
\left. \Gamma^{(n)}_{s\bar s A}(p) \right|_{0,0, 2} := \left.\Gamma^{(n)}_{l\bar l A}(p) \right|_{0,0, 2}\,,
\end{align}
as detailed in \cite{Gao:2020qsj}, Here $s$ is the strange quark and $l=u,d$ are the isopsin-symmetric light flavours. Then, $\Delta\Gamma^{(3)}_{s\bar s A}$ also comprises the differences of vacuum diagrams with light and strange quarks.

In contradistinction to the use of the difference-DSEs for gluon propagator and quark-gluon vertex, we solve the full quark gap equation for $\Gamma^{(2)}_{q\bar q}$. The set of DSEs considered in the present work is depicted in \fig{fig:gap} and \fig{fig:Vertices}, where we display the full DSEs also for gluon propagator and quark-gluon vertex instead of the  respective difference-DSE for the sake of simplicity. We also emphasise, that the computation of the quark propagator (and not its difference $\Delta\Gamma^{(2)}_{q\bar q}$ ) from the quark gap equation is the first consistency check of the procedure (and the input), as it can be compared with existing functional and lattice data for the vacuum quark propagator.

The difference-DSEs for the gluon propagator and quark-gluon vertex are solved within further truncations discussed below: \\[-1.5ex]

\textit{Difference-DSE for $\Delta\Gamma^{(2)}_{AA}$, see \fig{fig:gap}:} In short: We drop the ghost loop in the difference-DSE for the gluon propagator, as it gives negligible contributions. Furthermore, the full three-gluon vertex is approximated by its classical counterpart.

In detail: Ghost-gluon correlation functions show a negligible dependence on quark-content, temperature and density. Hence we can use their vacuum counterparts throughout. Then, the ghost loop in the difference-DSE for the gluon propagator only shows a thermal dependence, which we also drop, as it turns out to be negligible. We approximate the three-gluon vertex with its classical counterpart, as the three-gluon vertex dressing of the classical tensor structure only runs mildly for momenta $p^2\gtrsim 1$\,GeV. For smaller momenta it drops rapidly and even turns negative (but stays small). In this regime the strange and density contributions from the three-gluon diagram are negligible, as well as the subleading thermal fluctuations. This summarises the truncations used in the difference-DSE for the gluon propagator.  \\[-1.5ex]

\textit{Difference-DSE for $\Delta_{q\bar q A}^{(3)}$, see \fig{fig:Vertices}:} In short: We only compute the first two diagrams in the first line of \fig{fig:Vertices} within the difference-DSE for the quark-gluon vertex. We consider this approximation to be quantitatively small in all temperatures considered in the present work, and the density regime $\mu_B\lesssim 400$\, MeV.  For larger densities, its reliability decreases. There, the discussion of the systematic error is based on the comparison between  different functional approaches, fRG and DSE, used quite different resummations.

In detail: The third diagram in the first line (with the four-quark vertex) and the first diagram in the second line (with the quark-gluon scattering vertex) carry some information about hadron resonances. We consider in particular the neglect of the first diagram in the difference-DSE as the source for the largest systematic error in the present approximation: In the vacuum it is completely dominated by the  scalar-pseudoscalar tensor structure (sigma-pion resonance channel), the lowest lying hadronic states. This already follows from conceptual considerations and is at the root of e.g.\ chiral perturbation theory. It is corroborated by the $N_f=2$-flavour vacuum computation with the fRG in \cite{Cyrol:2017ewj} within a Fierz-complete computation with all tensor structures, which serves as the vacuum input here. At finite density we expect a change of the dominance order to a diquark channel, see in particular the analysis  \cite{Braun:2019aow} in QCD,  also \cite{Braun:2017srn, Braun:2018bik}. This change is also reflected in the non-trivial pion dispersion observed in \cite{Fu:2019hdw} at larger density and sizable quark condensate, see the red-hatched regime in \Fig{fig:finalphase}. We infer furthermore from DSE-studies with resonance contributions (hadronic and diquark resonances) in the  \cite{Eichmann:2015kfa, Fischer:2018sdj, Gunkel:2019xnh}, that the resonance-triggered deformations of the phase structure at finite density are subleading, though sizable. Consequently, in the ongoing quest for a systematic improvement towards quantitative reliability of functional QCD studies at large densities we will include this term in future work. This can be  done on the basis of the vacuum and finite density fRG-data in \cite{Cyrol:2017ewj, Fu:2019hdw}, as well as the difference-DSE for the four-quark vertex, also utilising bound-state approaches as put forward in \cite{Eichmann:2015kfa, Fischer:2018sdj, Gunkel:2019xnh}.

As already discussed for the difference-DSE for the gluon propagator, the contribution of the ghost diagram in the second line of \fig{fig:Vertices} to the difference-DSE of the quark-gluon vertex can safely be neglected. This also holds true for the contributions of the purely gluonic two-loop diagrams: the last diagram effectively dresses the classical three-gluon vertex in the first diagram in the second line, and this dressing is mild as discussed above. In turn, the penultimate diagram adds a tree-level contribution to the 1PI quark-gluon vertex. In combination this results in a diagram with the Bethe-Salpeter (BS) scattering vertex, that is relevant for the full consideration of resonances. This summarises the truncations used in the difference-DSE for the quark-gluon vertex.\\[-1.5ex]

\begin{figure}[t] 
	\includegraphics[width=1\columnwidth]{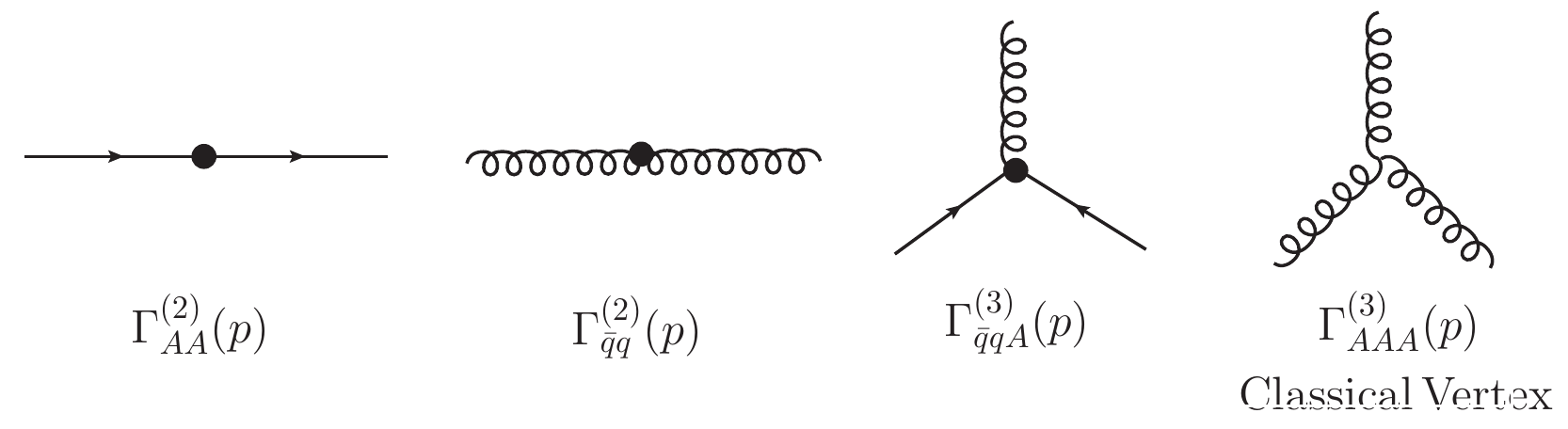}
	\vspace{-.1cm}
	
	\caption{Set of correlation functions, whose DSEs or Difference-DSEs are computed self-consistently in the present work: all quantum, temperature and density corrections computed here are fed back into the diagrams in the DSEs. We take into account all dressings of the quark-gluon vertex. For more details see Section~\ref{sec:truncation}.}\label{fig:Ingred}
\end{figure}
We conclude the discussion of the truncation with a short wrap-up: First we emphasise that the full system of DSEs, depicted in \fig{fig:gap} and \fig{fig:Vertices}, is computed self-consistently within the truncation described above: all quantum, temperature and density corrections computed here are fed back into the diagrams in the DSEs. The respective system of propagators and vertices is depicted in \fig{fig:Ingred}.

Secondly, on the basis of the detailed error analysis above we consider this truncation to be quantitative for $\mu_B\lesssim 400$\,MeV. In turn, for larger chemical potentials, the systematic error analysis based on the evaluation of the potential contributions of the dropped diagrams and tensor structures is not fully conclusive. In the present work we rely on a comparison with the fRG-results from \cite{Fu:2019hdw}. The existing and ongoing work on the resonance contributions in functional approaches (fRG and DSE-BSE), will lead to quantitative reliability of functional approaches in this regime within the next years.

\subsection{Gap equation and quark-gluon vertex DSE at finite temperature and density} \label{sec:quark+quark-gluon}

At finite temperature and density the quark two-point function reads,
\begin{align}\label{eq:quark2pointTmu}
\Gamma^{(2)}_{q\bar q}(\tilde p) = Z_q(\tilde p) \left[  \frac{Z^{\lt}_q(\tilde p)}{ Z_q(\tilde p) }\Bigl[  \gamma_0 \,\imag \tilde p_0 +  M_q(\tilde p)\Bigr]+
\vec \gamma \,\imag \vec p \right]\,,
\end{align}
with  $\tilde p= \left(p_0 - i \, \mu_B/3\,, \,{\bf p}\right)$, and quark Matsubara-frequencies $p_0 =2 \pi T\left(  n+\frac12\right)$. For more details see \cite{Gao:2020qsj}. The quark gap equation at finite temperature and density is given by
\begin{eqnarray}\nonumber
\Gamma^{(2)}_{q\bar q}(\tilde p)-S^{(2)}_{q\bar q}(\tilde p)
&=& \sumint\frac{dq_0}{2\pi} \! \int\frac{d^3{q}}{(2\pi)^3}\;   \Biggl[ G_{AA}{}_{\mu\nu}^{ab} (\tilde q+ \tilde p) \quad  \\[1ex]
& & \hspace{-1.5cm}\times \frac{\lambda^a}{2} {(-ig \gamma_{\mu})} G_{q\bar q}(\tilde{q})
\left[\Gamma^{(3)}_{q\bar qA}\right]^b_\nu (\tilde q,-\tilde p)\Biggr]\,.
\label{eq:DSEq}\end{eqnarray}
In \eq{eq:DSEq} $S^{(2)}_{q\bar q}$ is the inverse of classical quark propagator,  the full quark and gluon propagators are given by $G_{q\bar q} = (1/\Gamma^{(2)})_{q\bar q}$ and $G_{AA}= (1/\Gamma^{(2)})_{AA}$ respectively, and all  momenta are counted as incoming. Gluon (and ghost) Matsubara-frequencies are given by $q_0 + p_0 = 2 \pi T n$.

\begin{figure}[t] 
	\includegraphics[width=1.\columnwidth]{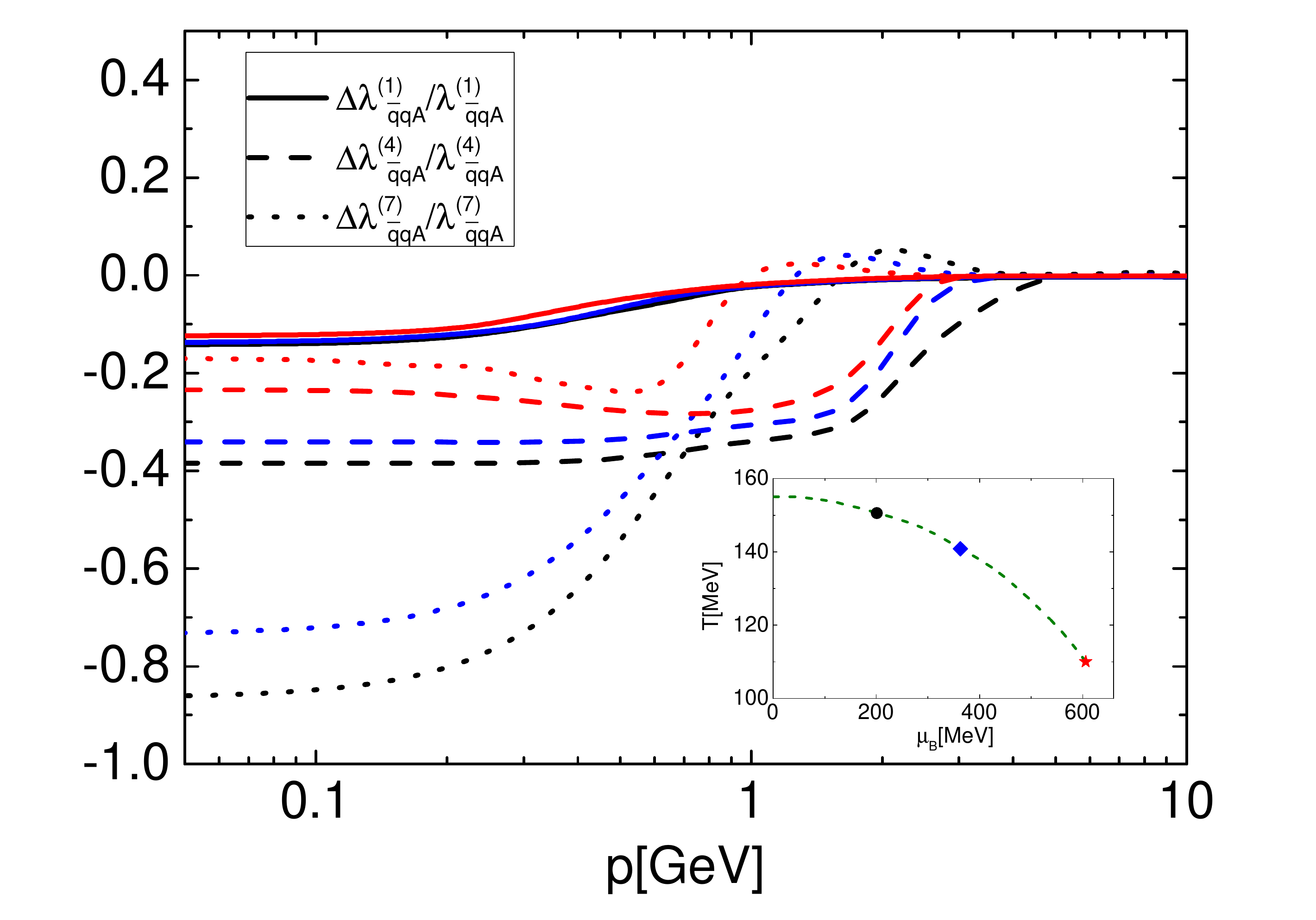}
	\vspace{-.4cm}
	\caption{The difference between finite $T,\,\mu$ and vacuum of the averaged  dressings for the three dominant structure,   $\lambda^{(1,4,7)}_{q\bar q A}$ of the tensors ${\cal T}^{(1,4,7)}_{q\bar q A}$ in  \eq{eq:TensorsQuarkGluon}, normalized by the respective vacuum dressing,  along the phase transition line with $(\mu_B, T)=(199, 150), (361, 140), (606, 110)$ MeV (\emph{Black, blue, red, respectively}). }\label{fig:Dressing}
\end{figure}
For the gluon propagator, we use the $O(4)$-symmetric approximation, in which the color-electric  and color-magnetic components agree.  For the quark-gluon vertex, the splitting is even weaker and we only keep the vacuum tensor structures. In \cite{Cyrol:2017ewj, Gao:2020qsj} the full transverse quark-gluon vertex has been written as follows,
\begin{align}\label{eq:FullGqbarqA}
\left[\Gamma^{(3)}_{q\bar q A}\!\right]^a_\mu (p, q)=\frac{\lambda^a}{2}\sum_{i=1}^8 \lambda^{(i)}_{q\bar q A}(p,q)\Pi^\bot_{\mu\nu}(k_+)\left[{\cal T}^{(i)}_{q\bar q A}\!\right]_\nu\! (p,q),
\end{align}
with the transverse  projection operator $\Pi^\trans_{\mu\nu}(k)=\, \delta_{\mu\nu}- k_\mu k_\nu/k^2$ and the Gell-Mann matrices $\lambda^a$. The tensor basis $\{{\cal T}_{q\bar q A}^{(i)}\}$ in \eq{eq:FullGqbarqA} is given by
\begin{align}
\begin{array}{lcl}
\left[{\cal T}^{(1)}_{q\bar q A}\right]^\mu(p,q) =-i \gamma^{\mu}\,, &\qquad&
\left[{\cal T}^{(5)}_{q\bar q A}\right]^\mu(p,q) =i{\kslash}_+ k_-^\mu\,,\\[2ex]
\left[{\cal T}^{(2)}_{q\bar q A}\right]^\mu(p,q) =k_-^\mu\,, &\qquad&
\left[{\cal T}^{(6)}_{q\bar q A}\right]^\mu(p,q)  = i\kslash_- k_-^\mu\,,\\[2ex]
\left[{\cal T}^{(3)}_{q\bar q A}\right]^\mu(p,q) = {\kslash_-}\gamma^\mu\,, &\qquad&
\left[{\cal T}^{(7)}_{q\bar q A}\right]^\mu(p,q) =\frac{i}{2} [\pslash,\qslash]\gamma^\mu\,, \\[2ex]
\left[{\cal T}^{(4)}_{q\bar q A}\right]^\mu(p,q) =\kslash_+\gamma^\mu\,, &\qquad&
\left[{\cal T}^{(8)}_{q\bar q A}\right]^\mu(p,q) = -\frac12[\pslash,\qslash] k_-^\mu\,.
\end{array}
\label{eq:TensorsQuarkGluon}\end{align}

In \cite{Gao:2020qsj} it has been shown, that the full results are already approached quantitatively by only considering the three dominant tensor structures, ${\cal T}^{(1,4,7)}_{q\bar q A}$. The vertex construction in \cite{Gao:2020qsj} for the thermal and density-fluctuations $\Delta\lambda^{(i)}_{q \bar q A}$ has utilised the STI and further gauge symmetry constraints. In turn, in the present work we compute the vertex dressings for $i=1,4,7$  directly with their respective (difference-)DSEs. The other vertex dressings are obtained with their respective symmetry constraints from the $\Delta\lambda^{(1,4,7)}_{q \bar q A}$, see \cite{Mitter:2014wpa, Cyrol:2017ewj, Gao:2020qsj},
\begin{align}
\begin{array}{rclcrcl}
\Delta\lambda^{(2)}_{q \bar q A}  &\approx & \frac{1}{2}\Delta\lambda^{(4)}_{q \bar q A}\,, &\qquad& \Delta\lambda^{(3,8)}_{q \bar q A}&\approx & 0\,, \\[2ex]
\Delta\lambda^{(5)}_{q \bar q A} &\approx & \frac{1}{2}\Delta\lambda^{(7)}_{q \bar q A}\,,&\qquad&   \Delta\lambda^{(6)}_{q \bar q A} &\approx & \frac{1}{12}\Delta\lambda^{(7)}_{q \bar q A}\,.
\end{array}
\end{align}
The results for $\Delta\lambda^{(1,4,7)}_{q \bar q A}$ along the phase transition line are displayed in \Fig{fig:Dressing}.

\section{Results}\label{sec:Results}

The functional fRG-DSE setup based on \cite{Gao:2020qsj} allows for a fully self-consistent computation of the QCD phase structure. The qualitatively novel feature  of the present computation is the self-consistent direct computation of the dressing of the three dominant tensor structures of the quark-gluon vertex at finite temperature and density, see  \Fig{fig:Dressing}. This qualitatively improves the reliability of the truncation at large densities as discussed in the previous Section~\ref{sec:FRG-DSE}.

\subsection{Renormalisation and benchmark results}

The current computation relies only on the determination of the fundamental parameters of QCD: the renormalisation scale of the present computation is $\mu=40$\,GeV   in accordance with the RG-scale of the fRG-input data. At this scale the strong coupling agrees with the fRG-coupling, $\alpha_s(\mu^2) = 0.188$. Furthermore, the light current quark masses are determined with the physical pion mass, $m_\pi(m_l)=140$\,MeV with $m_{u/d}=m_l=4.7$ MeV (isospin symmetric case). In the $N_f=2+1$ flavour case, the strange current quark mass is fixed with the ratio of $m_s^{0}/m_l^{0}\approx 27$. This fixes all physics parameters in the system. No further parameter is introduced, in particular the present computation does not rely on infrared phenomenological parameters. A full discussion of self-consistent RG-schemes for DSEs for given input data will be given in \cite{GPP:2020}.

\begin{figure}[t]
\vspace{-0.05cm}
	\includegraphics[width=1\columnwidth]{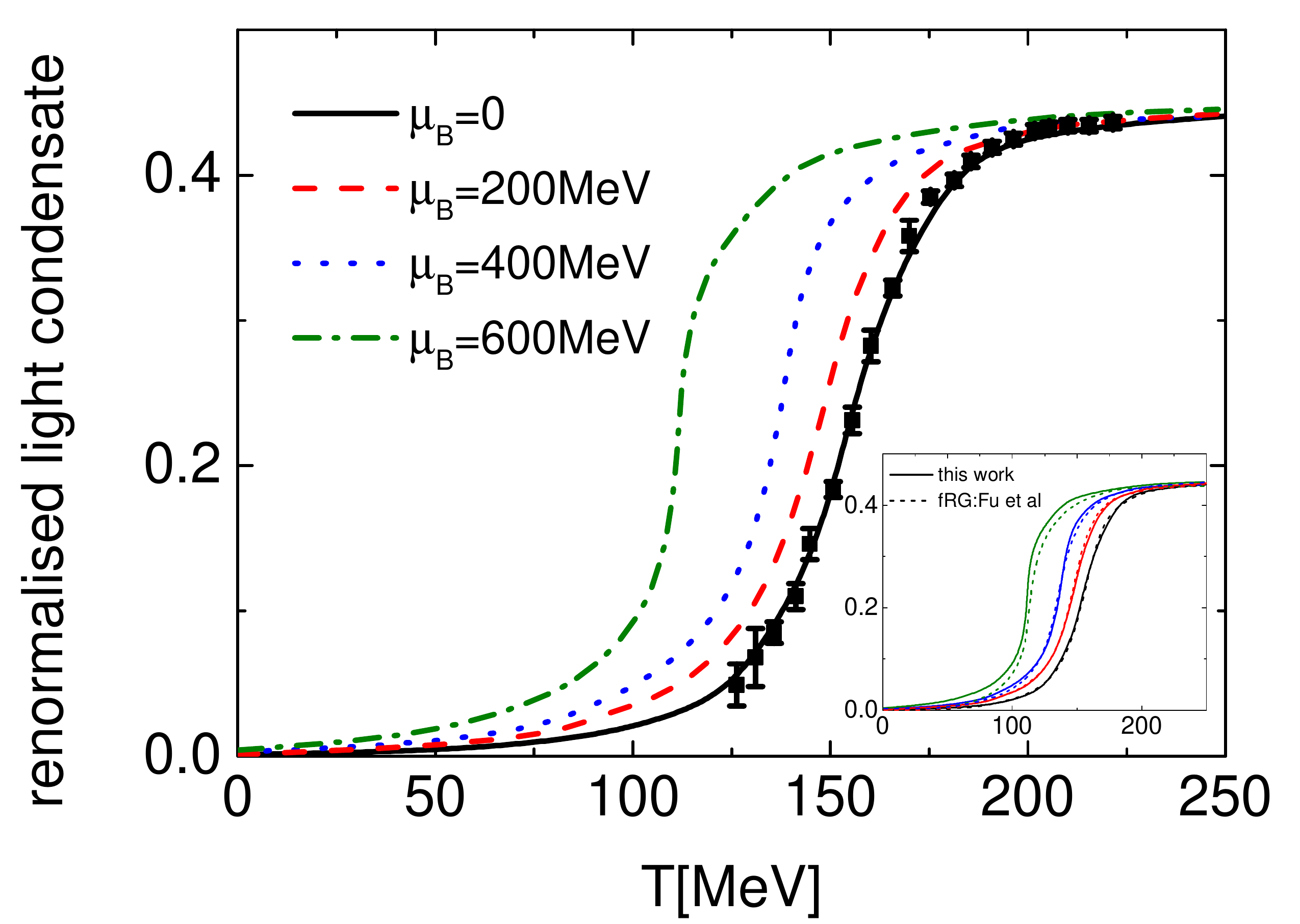}	
	\caption{The renormalised light chiral condensate at $\mu_B=0,\,200,\,400,\,600$ MeV  (\emph{Black, blue, red, olive  curves, respectively}) as a function of $T$. The result agrees with the lattice \cite{Borsanyi:2010bp} at vanishing chemical potential, and with the fRG-results from \cite{Fu:2019hdw} for all chemical potentials, see the inlay. }\label{fig:CondenMu}
	\vspace{-.2cm}
\end{figure}

The chiral transition temperature $T_c(\mu_B)$ is derived from the thermal susceptibility $\partial_T \Delta_{l,R}$ of the renormalised light chiral condensate $\Delta_{l,R}$ as in \cite{Fu:2019hdw,Gao:2020qsj},
\begin{align}\label{eq:chiralcondren}
	\Delta_{l,R} = \frac{1}{{2 \cal N}_R}\sum_{q=u,d}\Bigl[\Delta_{q}(T,\mu_B) -
	\Delta_{q}(0,0)\Bigr]\,.
\end{align}
where ${\cal N}_R$ is a convenient normalization which leads $\Delta_{l,R}$ dimensionless, and here we choose ${\cal N}_R=m^4_\pi$. The quark condensates $\Delta_{q_i}$ with $q_i=u,d,s$ are given by
\begin{align}\label{eq:chiralcondG}
	\Delta_{q_i}\simeq  - m_{q_i}^0
	T\sum_{n\in\mathbb{Z}} \int \frac{d^3 q}{(2 \pi)^3}
	\tr \,G_{q_i\bar q_i} (q)\,.
\end{align}
The results for \eq{eq:chiralcondren} at vanishing chemical potential are in quantitative agreement with the respective lattice results, see \fig{fig:CondenMu}. The chiral crossover temperature at vanishing density or chemical potential is readily derived from the susceptibility, and we obtain for $2$- and $2+1$-flavour QCD,
\begin{align}
\label{eq:Tcmu=0}
T_{c,N_f=2}= 166.0\,\textrm{MeV}\,,\quad T_{c,N_f=2+1}= 155.1\,\textrm{MeV}\,,
\end{align}
in quantitative agreement with the results obtained with the STI-vertices in \cite{Gao:2020qsj}:  $T_{c,N_f=2}= 166\,\textrm{MeV}$ and $T_{c,N_f=2+1}= 154\,\textrm{MeV}$. These results provides further support to the STI-construction of the thermal and density-corrections of the quark-gluon vertex put forward in \cite{Gao:2020qsj}. There, it has been carefully argued that such a construction is quantitatively reliable for the temperature regime of interest and not too large chemical potential.

A further benchmark computation concerns the curvature of the chiral phase boundary at vanishing chemical potential. It is determined from the transition temperature $T_c(\mu_B)$ within the expansion about $\mu_B=0$,
\begin{align}\label{eq:curv}
\frac{T_c(\mu_B)}{T_c}=1-\kappa\, \left(\frac{\mu_B}{T_c}\right)^2+\lambda\,\left(\frac{\mu_B}{T_c}\right)^4+\cdots\,,
\end{align}
with $T_c=T_c(\mu_B=0)$. This leads us to
\begin{align}\label{eq:kappaNf2-2+1}
\kappa_{N_f=2}=0.0175(7)\,,\qquad
\kappa_{N_f=2+1}=0.0147(5)\,.
\end{align}
Again, \eq{eq:kappaNf2-2+1} is in quantitative agreement (within the error bars) with the results obtained with the STI-vertices in \cite{Gao:2020qsj}:  $\kappa_{N_f=2}=0.0179(8)$ and $\kappa_{N_f=2+1}=0.0150(7)$, and provides further evidence for the quantitative reliability of the STI-construction for not too large densities. Respective results from the fRG, \cite{Fu:2019hdw,Braun:2020ada}, and lattice simulations, e.g.\ \cite{Borsanyi:2010bp,Bazavov:2017dus,Cheng:2007jq} also agree well with the results \eq{eq:Tcmu=0} and \eq{eq:kappaNf2-2+1}. For a comprehensive comparison see \cite{Fischer:2018sdj, Fu:2019hdw, Gao:2020qsj}.

\subsection{Chiral phase structure  and critical end point}

Now we proceed to the central result of the current work, the chiral phase structure at finite density. It is derived from the thermal susceptibility of the renormalised light chiral condensate \eq{eq:chiralcondren}, which is depicted as a function of temperature in \Fig{fig:CondenMu} for selected chemical potentials. As $\mu_B$ increases,  the crossover steepens, finally leading to 2nd order CEP and a first order regime.

\begin{figure}[t] 
	\vspace{-0.5cm}
	\includegraphics[width=0.98\columnwidth]{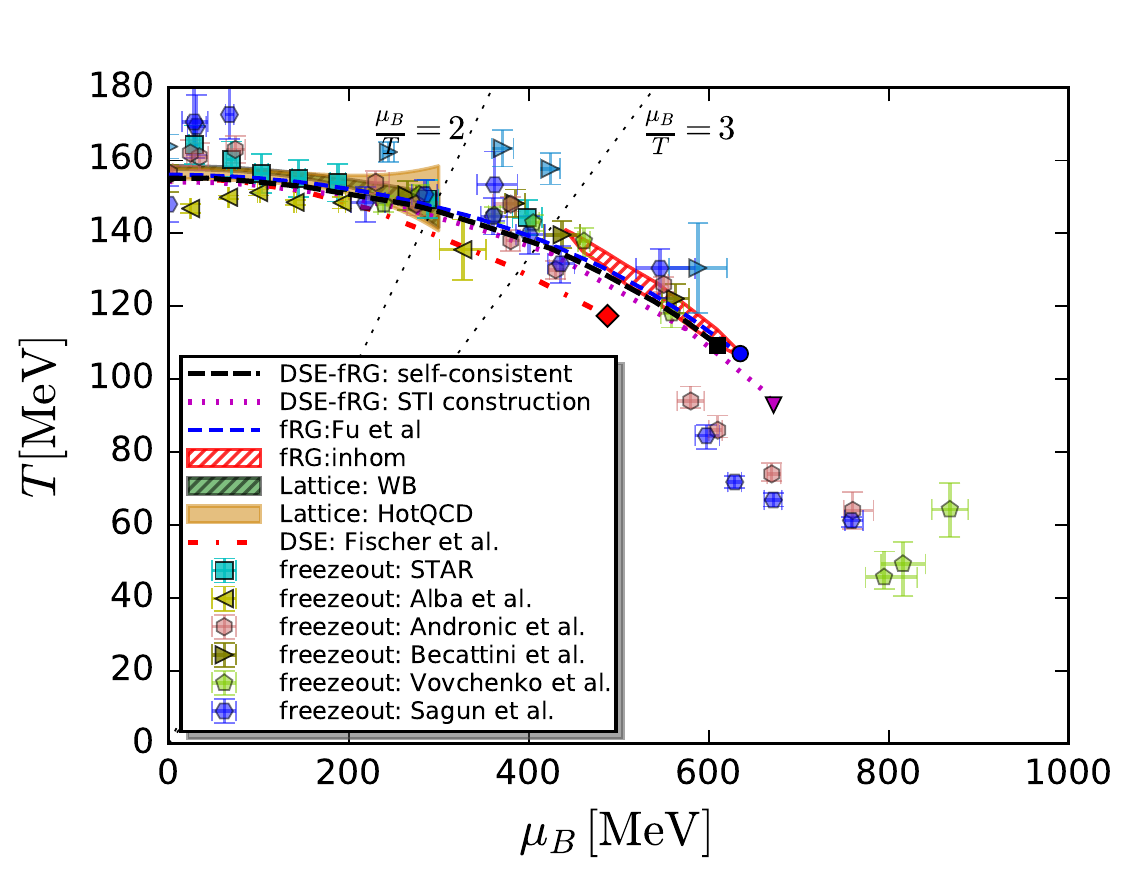}
	\caption{Phase diagram for 2+1 flavour QCD in comparison to other theoretical results and phenomenological freeze-out data. The blue dashed line displays the results from the current work. For small chemical potential the results agree well with the fRG and DSE, \cite{Fu:2019hdw, Fischer:2018sdj} and lattice results, \cite{Borsanyi:2020fev} (WB), \cite{Bazavov:2018mes} (HotQCD). In the red hatched regime the fRG-results from \cite{Fu:2019hdw} show a minimum of the pion dispersion at non-vanishing spatial momentum together with a sizable chiral condensate. This may indicate an inhomogeneous regime. For large chemical potential the results agree quantitatively with the fRG-results from \cite{Fu:2019hdw} including the location of the CEP. \\
		\textit{Freeze-out data}: \cite{Adamczyk:2017iwn} (STAR),
		\cite{Alba:2014eba} (Alba {\it et al.}), \cite{Andronic:2017pug}
		(Andronic {\it et al.}), \cite{Becattini:2016xct} (Becattini {\it et
			al.}), \cite{Vovchenko:2015idt} (Vovchenko {\it et al.}), and
		\cite{Sagun:2017eye} (Sagun {\it et al.}). Note that freeze-out data
		from Becattini {\it et al}.\ with (\textit{light blue}) and without (\textit{dark
		green}) afterburner-corrections are shown in two different
		colors.}\label{fig:finalphase}
\end{figure}

The resulting phase structure is depicted in \Fig{fig:finalphase}, together with the DSE-results with STI-vertices, \cite{Gao:2020qsj}, the fRG-results \cite{Fu:2019hdw}, and results from lattice simulations, \cite{Borsanyi:2020fev, Bazavov:2018mes}. For a comprehensive comparison see \cite{Fischer:2018sdj, Fu:2019hdw, Gao:2020qsj}.

In summary, the present results as well as that from \cite{Gao:2020qsj} agree with up-to-date functional results and lattice results for densities $\mu_B/T \lesssim 2$. For densities $\mu_B/T \gtrsim 2$ the present results agree with the fRG-results in \cite{Fu:2019hdw} including the location of the critical end point.

As discussed in detail in Section~\ref{sec:truncation}, for chemical potentials $\mu_B/T\gtrsim 3$,  the quantitative reliability of the approximations used in the respective functional approaches is successively lost, due to the lack of control of all potentially resonant quark-interactions and in particular diquark channels. This concerns the present fRG-assisted DSE-approach within the truncation used here, as well as the fRG computation used in \cite{Fu:2019hdw}. Still, the impressive agreement of the results from both approaches including the location of the CEP is a non-trivial reliability check. For $N_f=2+1$ flavours we get
\begin{align}\label{eq:CEP2+1}
	(T,{\mu_B})_{_{\tiny{\text{CEP}}}}
	=(109, 610)\,\textrm{MeV}\,,\quad
	&\  \frac{{\mu_B}_{_{\tiny{\text{CEP}}}}}{T_{_{\tiny{\text{CEP}}}}}
	= 5.59\,.
\end{align}
This result agrees quantitatively with the fRG-prediction $(T,{\mu_B})_{_{\tiny{\text{CEP}}}}=(107, 635)\,\textrm{MeV}$ with the ratio ${{\mu_B}_{_{\tiny{\text{CEP}}}}}/{T_{_{\tiny{\text{CEP}}}}}= 5.54$ in \cite{Fu:2019hdw}.

While this quantitative agreement between the fRG-computation and the present DSE-computation is indeed very intriguing and has to be seen as a respective support of the reliability of the results, it has to be taken with a grain of salt. As emphasised before, for quantitative predictions the approximations in both functions approaches have to be improved in order to allow for the potential resonance as well as competing order effects. Only then firm predictions on the location of the CEP can be made. Still, these improvements are within reach, and in summary the results of the present work together with the fRG-results of \cite{Fu:2019hdw} entail, that in functional approaches to QCD we are but one step away from quantitative statements concerning the phase structure at large densities.

\section{Summary}

In this work we have evaluated the QCD phase structure with a combination of the functional renormalisation group (fRG) and Dyson-Schwinger (DSE) equations: fRG-results from \cite{Cyrol:2017ewj} for the $2$-flavour gluon propagator and quark-gluon vertex in the vacuum have been used as an input for the DSEs of $2$- and $2+1$-flavour QCD at finite temperature and density. We have solved the coupled set of DSEs for the quark propagator and for the strange-quark, thermal and density corrections of the gluon propagator and quark-gluon vertex for $2$- and $2+1$-flavour QCD. No phenomenological infrared parameters have been tuned, and the only parameters are the fundamental ones in QCD: the current quark masses which are determined with the pion pole mass, and, in the 2+1 flavour case, also with the ratio of light and strange current quark masses.

After these successful benchmarks the setup has been used for the computation of the phase structure for $2$- and $2+1$-flavour QCD at finite temperature and density. For $2+1$-flavour QCD the results have been compared with other theoretical results as well as freeze out data, see \Fig{fig:finalphase} and the discussion in Section~\ref{sec:Results}. At chemical potentials $\mu_B/T\lesssim 2$ the present result agrees quantitatively with up-to-date functional and lattice results. For chemical potentials $\mu_B/T\gtrsim 2$, lattice simulations are obstructed by the sign problem. In this regime the present fRG-assisted DSE-results agree quantitatively with the fRG-results in \cite{Fu:2019hdw}, including the location of the critical end point. This non-trivial agreement enhances the reliability of the respective results, given the very different resummation schemes of DSE and fRG used in particular in the matter sector. The self-consistent computation of the quark gap equation and the thermal and density fluctuations of the gluon propagator and quark-gluon vertex leads to a critical endpoint at $(T_\textrm{\tiny{CEP}},{\mu_{B}}_{\textrm{\tiny{CEP}}})=(109,610)$\,MeV, see \eq{eq:CEP2+1}.

The present results and the analysis of the systematic error at large chemical potentials calls for a decisive final improvement of the truncation of functional approaches, both the fRG and the DSE, for full quantitative reliability in this regime: the inclusion of potentially dominant resonance structures of multi-quark states, in particular in the diquark channel, see \cite{Braun:2019aow}. This is subject of current work both in the fRG- and DSE-approach and we hope to report on results soon. \\[-2ex]

\noindent{\bf Acknowledgements}\\[1ex]
We thank J.~Braun, G.~Eichmann, C.~S.~Fischer, W.-j.~Fu, J.~Papavassiliou, F.Rennecke, B.-J.~Schaefer and N.~Wink for discussions. F.~Gao is supported by the Alexander von Humboldt foundation. This work is supported by EMMI and the BMBF grant 05P18VHFCA. It is part of and supported by the DFG Collaborative Research Centre SFB 1225 (ISOQUANT) and the DFG under Germany's Excellence Strategy EXC - 2181/1 - 390900948 (the Heidelberg Excellence Cluster STRUCTURES).

\bibliography{ref-lib}
\end{document}